\documentclass[aps,prb,twocolumn,showpacs,superscriptaddress]{revtex4}
\usepackage{graphicx}
\usepackage[colorlinks,citecolor=blue,linkcolor=blue]{hyperref}
\begin{document}

\title{Topological Magnon Insulator in Insulating Ferromagnet}
\author{Lifa~Zhang}
\email{phyzlf@gmail.com}
\affiliation{Department of Physics and Centre for Computational
Science and Engineering, National University of Singapore, Singapore
117542, Republic of Singapore }

\author{Jie Ren}
\email{renjie@lanl.gov}
\affiliation{Theoretical Division, Los Alamos National Laboratory, Los Alamos, New Mexico 87545, USA}

\author{Jian-Sheng~Wang}
\affiliation{Department of Physics and Centre for Computational
Science and Engineering, National University of Singapore, Singapore
117542, Republic of Singapore }
\author{Baowen~Li}
\affiliation{Department of Physics and Centre for Computational Science and Engineering, National University of Singapore, Singapore 117542, Republic of Singapore }
\affiliation{NUS Graduate School for Integrative Sciences and Engineering, Singapore 117456, Republic of Singapore}
\affiliation{NUS-Tongji Center for Phononics and Thermal Energy Science and Department of Physics, Tongji University, 200092 Shanghai, PR China}
\date{Mar 27, 2013}

\begin{abstract}
In the ferromagnetic insulator with the Dzyaloshinskii-Moriya interaction, we theoretically predict and numerically verify a topological magnon insulator, where the charge-free magnon is topologically protected for transporting along the edge/surface while it is insulating in the bulk. The chiral edge states form a connected loop as a $4\pi$- or $8\pi$-period
M\"{o}bius strip in the spin-wave vector space,
showing the nontrivial topology of magnonic bands.
Using the nonequilibrium Green's function method,
we explicitly demonstrate that the one-way chiral edge transport is indeed topologically protected from defects or disorders. Moreover, we show that the topological edge state mainly localizes around edges and leaks into the bulk with oscillatory decay.
Although the chiral edge magnons and energy current prefer to travel along one edge from the hot region to the cold one, the anomalous transports are identified in the opposite edge, which reversely flow from the cold region to the hot one. Our findings could be validated within wide energy ranges in various magnonic crystals, such as Lu$_2$V$_2$O$_7$.


\end{abstract}


\maketitle
\section{Introduction}
Topological insulator, as a novel state of quantum matter,
is characterized by an insulating bulk band gap and conducting
gapless edge/surface states protected by symmetries
\cite{hasan10,qi11}. It has been theoretically predicted and experimentally observed in a variety of systems and becomes a hot spot because of its theoretical importance in condensed matter physics and wide potential applications in dissipationless spin-based electronics
(spintronics) \cite{spin1}. However, due to the fact that the spin transport in
topological insulators is carried by electrons, dissipations can
not be really avoided.

Magnon Hall effect, as a consequence of the Dzyaloshinskii-Moriya (DM) interaction \cite{dzya58,moriya60} that plays a role of vector potential similar to the Lorentz force, has been predicted and observed in magnetic insulators \cite{katsura10,onose10,matsumoto11}. Compared with spin current, where the dissipation from Joule heating is still inevitable due to the electronic carriers, the magnon Hall effect is more promising in device applications because of the long-range coherence of charge-free spin wave \cite{hirsch99,murakami03,sinova04}. Magnons are collective excitations of localized spins in a crystal lattice and
can be viewed as quantized quasiparticles of spin waves. Recently,
magnon excitation \cite{serga04,demidov09}, localization \cite{jorzick02} and interference \cite{podbielski06} have been experimentally realized. 
The technical advancements offer the perspective of various magnonic devices,
 and a new discipline -- magnonics -- has emerged and is growing exponentially  \cite{kruglyak06,neusser09,kruglyak10,serga10,lenk11}.
The charge-free property of magnon makes it promising to achieve dissipationless transport and control in insulating magnets.

\begin{figure}
\includegraphics[width=1.00\columnwidth]{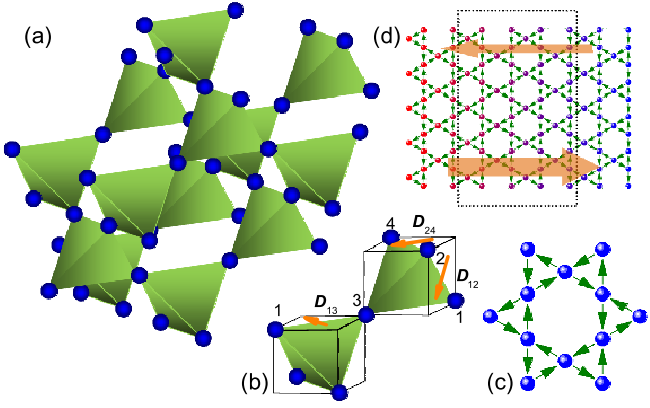}%
\vspace{-3mm}
\caption{\label{fig1_lattice} (color online). (a) Pyrochlore
crystallographic structure of the sublattice of magnetic atoms V
of  Lu$_2$V$_2$O$_7$. (b) Two tetrahedrons in the pyrochlore
lattice, where DM vectors on bonds 1-3, 1-2, and 2-4 are indicated by
orange arrows. (c) Schematic magnetic flux due to DM interaction
in the [111] plane of the pyrochlore lattice, i.e., a kagome
lattice. The coupling of two sites along the arrows is
$(J+iD)S$ while the opposite direction corresponds to $(J-iD)S$.
(d) The quasi-one-dimensional kagome-lattice strip. The area
enclosed by the dotted line can be regarded as a center which is
connected to two semi-infinite leads in equilibrium at
temperatures $T_L$ (left) and $T_R$ (right). The two big arrows
schematically depict the magnitudes and directions of energy flows along the lattice edges when $T_L>T_R$.  The width of the strip example is
$W=5$, which is defined as the number of atoms in the left column of each unit cell.
}
\end{figure}

Therefore, it will be of great general interest for both theorists
and experimentalists that we find in this work a new intriguing
quantum state that magnon while insulated in
the bulk, can nondissipatively transport along edges/surfaces in
the absence of backscattering from defects and disorders
due to the nontrivial topology of magnon's band structures.
We name this novel state
topological magnon insulator (TMI) and believe that due to the
robust dissipationless magnon transport, the TMI in insulating
magnets could provide widely potential applications in
nondissipative magnonics and micro-spintronics.

\section{Spin-wave Hamiltonian}
The magnon Hall effect was experimentally observed in insulating ferromagnet Lu$_2$V$_2$O$_7$ \cite{onose10} with a pyrochlore lattice, in which
the magnetic atom vanadium has a corner-sharing-tetrahedra
sublattice, that is, a stacking of alternating  kagome and
triangular lattices along the [111] direction, as shown in
Fig.~\ref{fig1_lattice} (a). To study magnon transport in the
ferromagnetic insulator, the Hamiltonian can be written
as \cite{dzya58,moriya60,bose94}:
\begin{equation}\label{eq_ham} \mathcal{H}=
\sum\limits_{\left\langle {mn} \right\rangle } {[ - J \vec S_m
\cdot \vec S_n  + \vec D_{mn}\cdot(\vec S_m  \times \vec S_n )]} -
g\mu _B \vec H_0 \cdot\sum\limits_n {\vec S_n },
\end{equation}
where $\vec S_n$ is the spin angular momentum at site $n$; $-J$ denotes
the nearest-neighbor coupling; $\vec D_{mn}$ is the DM interaction
between site $m$ and $n$; the last term comes from the Zeeman effect under an external field $\vec H_0$.

As shown in Fig.~\ref{fig1_lattice} (b), in a single tetrahedron,
the DM vector is perpendicular to the corresponding bond and
parallel to the surface of the surrounding cube \cite{onose10,elhajal05,kotov05}. Since the component of $\vec D_{mn}$ perpendicular to $\vec{z}=\vec H_0/H_0$ does not contribute to the Hamiltonian (\ref{eq_ham}) up to quadratic order of the deviation of $\vec S$ \cite{onose10}, we only retain $D_{mn}=\vec D_{mn} \cdot \vec{z}$. When applying a magnetic field along $\vec{z}=[111]$ direction, all the projections of the DM interaction between inter-layer sites $m$ and $n$ are zero ($D_{13}=D_{23}=D_{43}=0$); and all the ones for inner-layer sites are nonzero and equal
($D_{12}=D_{24}=D_{41}=D$). Therefore, with the magnetic field along $[111]$ direction, 
the kagome sublattice structure will
play a key role for the presence of TMI effect in
Lu$_2$V$_2$O$_7$. In addition, the two-dimensional kagome lattice sheet could be obtained through doping one-quarter of the sites [e.g., site 3 in Figure~\ref{fig1_lattice} (b)] of a pyrochlore lattice by nonmagnetic atoms \cite{shores05,olariu08,colman08}.

In the following we first discuss a general
kagome lattice with DM interaction; later we will
incorporate actual parameters of a thin film of
Lu$_2$V$_2$O$_7$ with a kagome layer of vanadium sublattice.
Using the relation of $
S^x  = {1 \over {2}}(S^ +   + S^ -  )$ and $S^y  = {1 \over {2i}}(S^ +   - S^ -  )
$, we can rewrite the Hamiltonian (\ref{eq_ham}) on a kagome lattice as
\begin{eqnarray}
\mathcal{H} = &-& \sum_{\langle {mn} \rangle } {\left( {{J + i D } \over 2}S_m^- S_n^+   + {{J - i D} \over 2}S_m^+ S_n^- \right)}  \nonumber \\
&-& \sum_{\langle {mn} \rangle } { J S_m^z S_n^z }  - h\sum\limits_n {S_n^z },
\end{eqnarray}
where $h=g\mu _B H_0$.
Now, applying the standard Holstein-Primakoff transformation \cite{holstein40,zhang08}, one can straightforwardly obtain the quadratic spin-wave Hamiltonian:
\begin{equation}
\mathcal{H}  = \sum\limits_{mn}{b_m^+ H_{mn} b_n} +E_0,
\label{eq_hamsw}
\end{equation}
where $b^+$ ($b$) denotes the operator raising (lowering) the spin component along  $\vec{z}$ direction. $E_0 = - JS^2\sum_n
{M_n/2}-NhS$ is the ground-state energy with $N$ the total number of sites and $M_n$ the number of nearest neighbors of the site $n$.
$H_{mn}=H_{nm}^*=(J\pm i D)S$ and $H_{nn}=JS M_n+ hS$. Figure \ref{fig1_lattice} (c) illustrates the direction of the DM interaction vector, that is,
the coupling between two sites along the direction of that arrow
corresponds to $(J+iD)S$, while the coupling of the opposite direction corresponds to
$(J-iD)S$. Due to the different types of loops in a unit cell of the kagome lattice, the DM interaction avoids cancellation thus induces the Hall effect \cite{katsura10}. As a consequence, the preserved DM interaction acts as a vector potential for the propagation of magnons similar to the magnetic field for the propagation of electrons, which is crucial for the manifestation of TMI effect.
The magnetic field decides the direction of spins at the ground state, and the induced Zeeman effect term just shifts the dispersion relation. We set magnetic field $H_0=0^+$ in the part of theoretical model calculations, and will input finite $H_0$ in the part for real-material calculations. Except in Sec.~\ref{secreal}, dimensionless units and $S=1/2$, $J=1$ are used without loss of generality.

\begin{figure}
\includegraphics[width=1.00\columnwidth,angle=0]{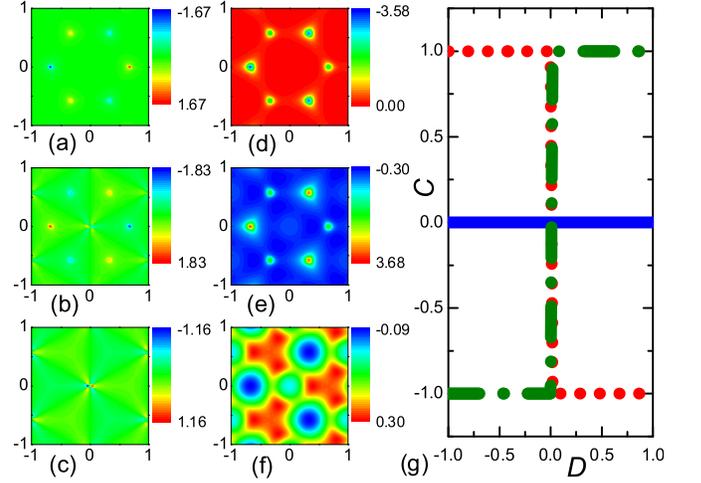}%
\caption{\label{fig2_chern} (a), (b) and (c) The Berry curvature of the three bands at zero DM interaction;  (d), (e) and (f) The Berry curvature of the three bands at nonzero DM interaction (D=0.2).  For all the insets (a)-(f), the horizontal and vertical axes correspond to wave vector $k_x$ and $k_y$, respectively; the unit is $2\pi/a$, where $a$ is the lattice spacing.  (g) The Chern numbers of the three energy bands for the two-dimensional periodic kagome lattice. The dotted, solid, and dashed lines correspond to Chern numbers of the first, the second, and the third bands, respectively.}
\end{figure}
\section{Chern numbers of bulk states}
The Eq.~(\ref{eq_hamsw}) resembles the tight-binding model, and each unit cell has three sites. For a two-dimensional periodic kagome spin lattice, we can perform the Fourier transformation as
\begin{equation}
b_{\vec R_l  + \vec r_m }  = \frac{1}{N_u}\sum\limits_{\vec k} {e^{ - i\vec k(\vec R_l  + \vec r_m )} } b_m (\vec k).
\end{equation}
Here, $N_u$ is the number of unit cells. $\vec R_l+ \vec r_m$ is the position of the $m$-th site in the $l$-th unit cell. Thus the spin-wave Hamiltonian can be written in the momentum space.

\begin{figure}
\includegraphics[width=1.00\columnwidth,angle=0]{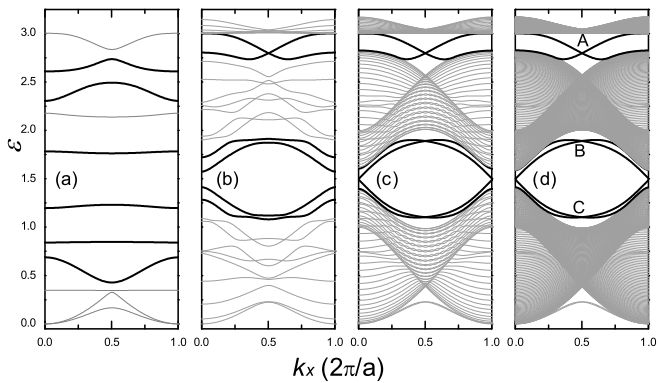}%
\caption{\label{fig3_edgesize} The dispersion relations of the periodic kagome strip lattices with different width sizes. The insets (a), (b), (c), and (d) correspond to $W=2$, $W=5$, $W=20$, and $W=80$, respectively. }
\end{figure}
Following the standard method to calculate the Berry phase \cite{berry84,xiao10,zhang11}, we can obtain the Berry curvature of the $n$-th band as:
\begin{equation}
B _{k_x k_y }^n  =  i\sum\limits_{n' \ne n } {\frac{{\varphi _n ^{\dag} \frac{{\partial H_{\rm{SW}} }}
{{\partial k_x }}\varphi  _{n '}\varphi _{n'} ^{\dag} \frac{{\partial H_{\rm{SW}} }}
{{\partial k_y }}\varphi _n   - (k_x  \leftrightarrow k_y )}}
{{(\varepsilon_n   -\varepsilon_{n '} )^2 }}}.
\end{equation}
Here $\varepsilon_n$ and $\varphi _n$ are the eigenvalue and eigenvector of the spin-wave Hamiltonian. The associated topological Chern number is obtained through integrating the Berry curvature over
the first Brillouin zone as
\begin{equation}\label{eq-cherni}
C^n   = \frac{1}
{{2\pi }}\int_{BZ} {dk_x dk_y B_{k_x k_y }^n  }.
\end{equation}
If the DM interaction is zero, the Berry curvatures of the three bands are shown in Fig.~\ref{fig2_chern}(a), (b) and (c): the maximum points have the opposite values; the sum of the Berry curvatures are zero, that is, the Chern numbers are zero at zero DM interaction as shown in Fig.~\ref{fig2_chern}(g). Therefore, the magnon Hall effect and topological magnon insulator effect is absent. If the DM interaction is nonzero, the Berry curvatures change dramatically and can not cancel each other. As shown in Fig.~\ref{fig2_chern}(d), in the whole momentum space, the Berry curvature of the first band is positive, which corresponds to the Chern number with the value of 1 as shown in  Fig.~\ref{fig2_chern}(g). And the Berry curvatures of the third band shown in Fig.~\ref{fig2_chern}(f) also can not cancel each other and the Chern number is $-1$. The Berry curvature of the second band also changes, but the Chern number keeps zero. The topological magnon insulator is only possible when the DM interaction is nonzero.

\section{Finite size effect for the dispersion relation of quasi-1D kagome lattice}
According to the spin wave Hamiltonian Eq.~(\ref{eq_hamsw}), we can calculate the dispersion relation ($\varepsilon$ vs $k_x$) of the quasi-1D periodic lattice, as shown in the Fig. 1 (d). In this figure, the left most column has 5 sites, thus we denote the width as $W=5$, in each unit cell of the quasi-1D kagome lattice there are $6W-1$ sites. As shown in Fig.~\ref{fig3_edgesize}, with increasing strip width, more modes appear in the energy bands;  the edge states will be gradually fixed and independent of size \cite{hatsugai93a}. If the width $W\geq20$, we find that the states in the bulk gap, that is, the edge states tend to be fixed, and in the bulk energy bands there are more and more branches. From $W=20$ to $W=80$, the edge states almost have no changes.

\begin{figure}
\includegraphics[width=1.00\columnwidth,angle=0]{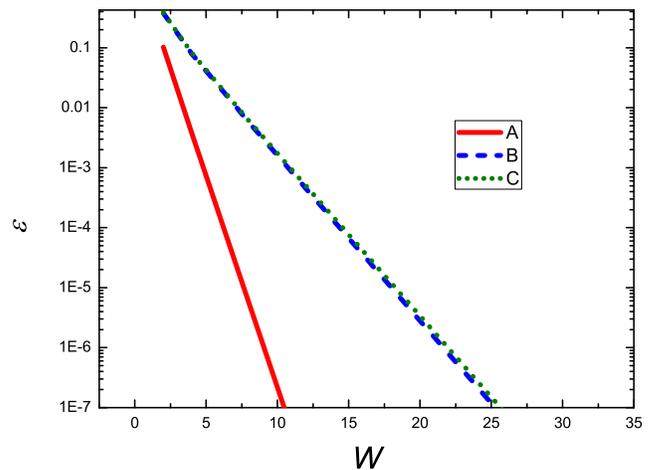}%
\caption{\label{fig4_degen} The energy differences vs the width of the periodic kagome strip at the anti-crossing points. The solid, dashed, and dotted lines correspond to energy differences at anti-crossing points A, B, and C, respectively in Fig.~\ref{fig3_edgesize} (d). }
\end{figure}

In the bulk energy gaps, we find the edge states have the trend to touch each other at the points $A, B, C$ as shown in Fig.~\ref{fig3_edgesize} (d). The energy difference of the corresponding edge states at the points $A, B, C$ is shown in Fig.~\ref{fig4_degen}. As the width increases, the energy difference decreases exponentially. If the system width is finite, the states in two edges have  nearly equal energy and momentum near the anticrossing points $A, B, C$; thus they can couple together to open an energy gap which decays exponentially with width increasing \cite{zhou08}. When the strip width increases to infinity, the edges are separated too far to interact with each other; thus they could have degeneracy in the dispersion relation. As shown in Fig.~\ref{fig4_degen}, the gap at point $A$ decays faster than that at $B$ and $C$, thus we can find the crossing at $A$ in the upper gap earlier than that in the lower gap. And after $W\geq20$, the energy differences at all the three points $A, B, C$ are very small, therefore it is very reasonable that we use $W=80$ in all the following numerical calculations to study the chiral edge state transport in the quasi-1D lattice with large-enough width.

\section{Topological magnon edge modes}
\begin{figure}
\includegraphics[width=1.00\columnwidth]{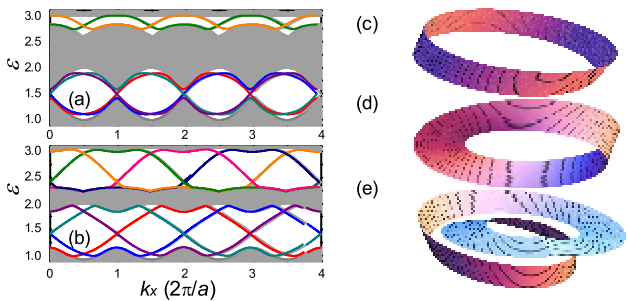}%
\vspace{-3mm}
\caption{\label{fig5_edge} (color online). The dispersion relation of chiral magnonic edge modes with non-zero DM interactions. (a) and (b) are dispersion
relations in range of $k_x\in[0, 8\pi/a]$ for $D/J=0.1$ and 0.4,
respectively. (c) A conventional cylinder strip with two
boundaries, of which each has a period of $2\pi$. (d) A M\"{o}bius strip
which only has one boundary of $4\pi$ period. (e) A looped
M\"{o}bius strip which only has one boundary of $8\pi$ period.}
\end{figure}

Because of the DM interaction, two edge
states within both energy gaps are twisted so that for each state $\varepsilon(\pi/a-k_x)\neq
\varepsilon(\pi/a+k_x)$ and they cross at $k_x=\pi/a$ in the first Brillouin zone, where $a$ is the lattice spacing. As shown in Fig.~\ref{fig5_edge}(a) for the case of $D/J=0.1$, in the upper bulk band gap, the two edge modes form a
continuous state with a period of $4\pi$, which can not be
disturbed to open a gap by weak disorders so that edge modes are topologically protected. However, when DM interaction is zero, the two edge states are easy to be perturbed to separate and open a gap, because they do not cross each other although they degenerate. In the lower bulk band gap, we find that four edge states will contribute to magnon transport within the energy gap.
When $D/J\approx0.4$ or larger, in both energy gaps there are four edge states [see Fig.~\ref{fig5_edge}(b)]. In a period of $8\pi$, any two of the four edge states have degeneracies and cross each other at  different points in the momentum space. All the four edge states form a continuous state with a period of $8\pi$ to transport magnons along two edges of the strip.

We can understand the topology of the edge states as follows. With zero DM interaction, the edge states are similar to the two boundaries of the conventional cylindric strip as shown in  Fig.~\ref{fig5_edge} (c), both of which have a period of $2\pi$ in Brillouin zone and transport separately along two edges. Due to the nonzero DM interaction, two edge states are twisted so that they cross each other and go into the other energy band after $2\pi$ in momentum space, thus form a closed loop with a period of $4\pi$. These edge states are similar to the one-sided M\"{o}bius strip with only one boundary, as shown in Fig.~\ref{fig5_edge} (d), where a line drawn starting from a point at the boundary will meet back at the ``other side'' after a circle of $2\pi$, then go back to the original point after a whole period of $4\pi$. The two edge states form one closed loop winding the bulk energy gap between the two bands, which are thus topologically protected from distortions. At larger DM interaction, four edge states contribute to the transport in the bulk gap, cross each other, and connect to form a closed $8\pi$-period loop which can be interpreted as the only one boundary of a looped M\"{o}bius strip as shown in Fig.~\ref{fig5_edge} (e). This looped M\"{o}bius strip also has only one boundary winding around the strip surface, thereby having the same topology as that of the conventional M\"{o}bius strip shown in Fig.~\ref{fig5_edge} (d).

The topological chiral edge state is related to the band topology characterized by Chern numbers of the bulk states \cite{hatsugai93a,qi06,scarola07,yao09,zhang11}, as shown in Fig. \ref{fig2_chern}(b). Since there are three sites in each unit cell, the two-dimensional infinite kagome lattice with Hamiltonian (\ref{eq_hamsw}) has three bands. When the DM interaction is absent, all the Chern numbers of three bands are zero so that there is no TMI effect. Accordingly, the winding numbers of edge states are zero as well, thus they are not topologically protected. With nonzero DM interaction, the Chern numbers of the lowest and highest energy bands become $\pm1$ that indicate the nontrivial topology; the one of the middle energy band is still zero. According to the relation between the Chern number and the winding number \cite{hatsugai93a}, the winding numbers of edge states in both bulk gaps have the same value of 1 or $-1$, which is consistent with the only one closed loop winding the bulk gap regardless of the period of $4\pi$, $8\pi$ or others.

\begin{figure}
\includegraphics[width=1.00\columnwidth]{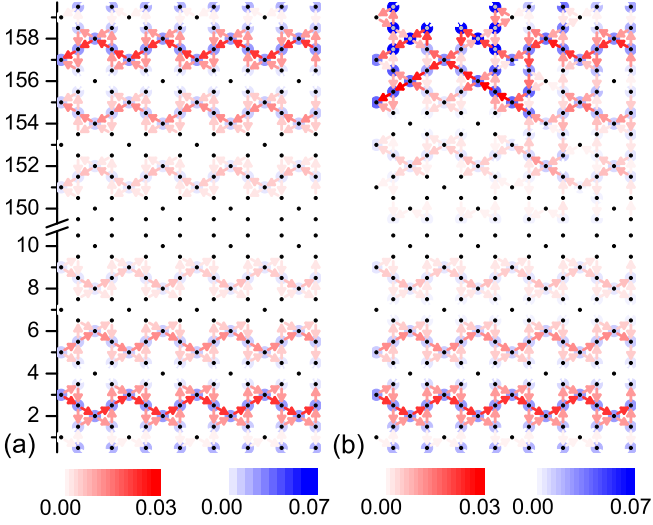}%
\caption{\label{fig6_localeq} (color online). The local energy current and density of state for edge magnon transport at equilibrium. (a) The uniform kagome lattice. (b) The lattice with a defect at the upmost site of the left fifth column. The red arrows, the blue dots, and the small black dots correspond to the local energy current, the local density of magnon, and atom sites, respectively. The color of the arrows and dots indicate the magnitude of the local current and density of states, respectively . Parameters are $\varepsilon=1.5$, $T_L=T_R=1.0$, $D/J=0.1$, and $a=1$.}
\end{figure}
\section{Topological magnon transport: the NEGF method}

To intuitively illustrate the topological magnon transport carried by chiral edge states, we choose some unit cells of the kagome lattice strip as a center region and set the rest as two semi-infinite leads in equilibrium at temperatures $T_L$ and $T_R$, respectively [see Fig.~\ref{fig1_lattice} (d)]. We then apply the nonequilibrium Green's function method \cite{haug96} to calculate the local density of magnons and the local energy current density of magnons. For the nonequilibrium magnon transport in such system, the Hamiltonian can be written as follows
\begin{equation}
\mathcal{H}  = \sum\limits{\mathcal{H}_\alpha  }  + \Bigl(\sum\limits_{lm}( {b_l^{L + }
H_{lm}^{LC} b_m^C  + b_m^{C + } H_{ml}^{CR} b_l^R } ) + {\rm h.c.}
\Bigr),
\end{equation}
where $\mathcal{H}_\alpha = \sum\limits_{lm} {b_l^{\alpha  + } H_{lm}^\alpha
b_m^\alpha},\alpha  = L,C,R$, here `$L,C,R$' denote the left lead,
the center part and the right lead, respectively.
The Hamiltonian matrix  of
the full system is
\begin{equation}\label{} H= \left( {\begin{array}{*{20}c}
   {H_L } & {H_{LC} } & 0  \\
   {H_{CL} } & {H_C } & {H_{CR} }  \\
   0 & {H_{RC} } & {H_R }  \\
\end{array}} \right).
\end{equation}
The retarded Green's function is defined as
\begin{equation}\label{}
G^r (t,t') =  - i\theta (t - t') \langle [b(t),b^ +  (t')]\rangle,
\end{equation}
where we set $\hbar=1$ for notational simplicity. In
nonequilibrium steady states, the Green's function is
time-translationally invariant so that it depends only on the
difference in time. Thus, the Fourier transform of $G^r (t-t')=G^r
(t,t')$ is obtained as
\begin{equation}\label{}
G^r[\varepsilon]=\int_{-\infty}^{+\infty}G^r (t)e^{i\varepsilon t}dt .
\end{equation}
Without interaction, the free Green's functions for three parts in
equilibrium can be written as:
\begin{equation}
\begin{array}{l}
 ((\varepsilon  + i\eta ) - H_\alpha  )g_\alpha ^r [\varepsilon ] = I,\quad
\alpha  = L,C,R, \\
 g_\alpha ^a [\varepsilon ] =g_\alpha^r[\varepsilon]^\dagger.
  \end{array}
\end{equation}
And there is an additional equation relating $g^r$ and $g^<$:
\begin{equation}
 g_{\alpha}^ <  [\varepsilon ] = f_{\alpha}(\varepsilon )[g_{\alpha}^a [\varepsilon ] - g_{\alpha}^r [\varepsilon ]],
\end{equation}
where $ f_{\alpha}(\varepsilon ) =  \langle b^ +  b \rangle  = [e^{\varepsilon /T_{\alpha}} -
1]^{-1} $ is the Bose-Einstein distribution function at the ${\alpha}$ part with
temperature $T_a$; we have set $k_B=1$.

For the quadratic Hamiltonian, the magnon transport is ballistic. The lesser Green's function can be solved as
\begin{equation}
 G^ <  [\varepsilon ] = G^r  [\varepsilon ] \Sigma^< [\varepsilon ]  G^a [\varepsilon]
 \end{equation}
where $G^a=(G^r)^\dag$ and the self energy
\begin{equation}
\Sigma^{r,<} [\varepsilon ]= H_{CL} g_L^{r,<}  [\varepsilon ]H_{LC}+ H_{CR} g_R^ {r,<}  [\varepsilon ]H_{RC}.
\end{equation}
The retarded Green's function has the same form as for the electron case
\begin{equation}
 G^r [\varepsilon] = \Bigl[\varepsilon + i\eta  - H_C  - \Sigma^r [\varepsilon ] ]^{ - 1}.
\end{equation}
The local density of magnon is given by \cite{zhang07}
\begin{equation}
\rho _n = {{i\hbar G_{nn}^ <  (\varepsilon )} \over {\pi a}}.
\end{equation}
The local energy current is given by \cite{jing09}
\begin{equation}\label{seq_localj}
\mathfrak{j}_{mn} (\varepsilon ) = {\varepsilon \over {2\pi }}{\mathop{\rm Re}\nolimits} [G_{mn}^ <  [\varepsilon ] H_{nm}  -  G_{nm}^ <  [\varepsilon ]H_{mn}].
\end{equation}
At the interface between the left lead and the center part, it reads
\begin{equation}
\mathfrak{j}_{mn} (\varepsilon ) = {\varepsilon \over {2\pi }}{\mathop{\rm Re}\nolimits} [G_{mn}^{CL, <}  [\varepsilon ] H_{nm}^{LC} -  G_{nm}^{ LC, <}  [\varepsilon ]H_{mn}^{CL}].
\end{equation}
Taking the trace to sum over all the local current in the interface, and integrating over all the energy, we then get the Landauer-like formula as
\begin{eqnarray}
 J &=&\int_0^\infty  {\sum\limits_{} {j_{mn} } } d\varepsilon\nonumber \\
 &=& \int_0^\infty d\varepsilon {\varepsilon \over {2\pi }} {\rm{Tr}} \bigl \{ {\mathop{\rm Re}\nolimits} \bigl( G^{CL, <}  [\varepsilon ] H^{LC} -  G^{ LC, <} [\varepsilon ]H^{CL} \bigr )\bigr \} \nonumber \\
 &=&\frac{1}{{2\pi }}\int_{0}^\infty
{\varepsilon \;[f_L (\varepsilon ) - f_R (\varepsilon)] T[\varepsilon ]} d \varepsilon
\end{eqnarray}
where the transmission is
\begin{equation}
T[\varepsilon ] = {\rm{Tr}} \bigl \{ G^r [\varepsilon ] \Gamma _L [\varepsilon ] G^a [\varepsilon ] \Gamma_R[\varepsilon ] \bigr \}.
\end{equation}
with the $\Gamma _\alpha $ functions given by $\Gamma _\alpha
= i(\Sigma _\alpha ^r  - \Sigma _\alpha ^a ).$

\begin{figure}[t]
\includegraphics[width=1.0\columnwidth]{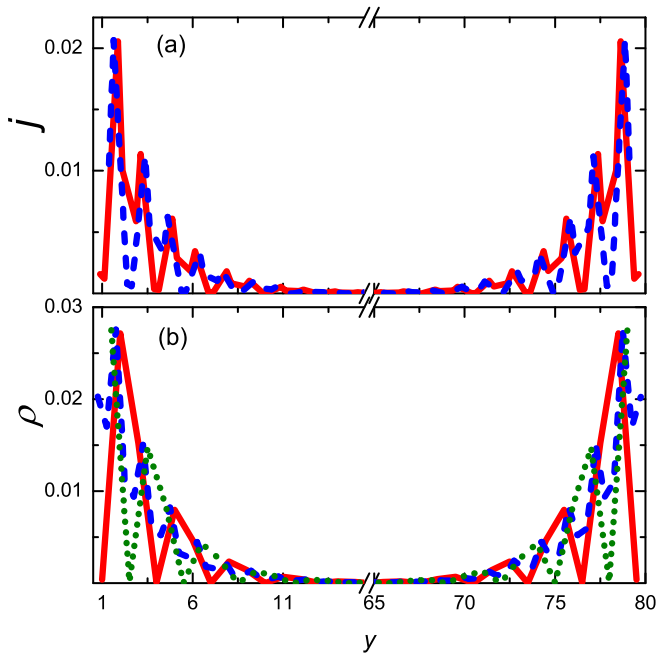}%
\caption{\label{fig7_jroy} (a) Local energy current vs the coordinate along the $y$ direction. The solid and dashed lines correspond to the local currents in two different columns in one unit cell. (b) Local density of states of magnons vs the coordinate along the $y$ direction. The solid, dashed, and dotted  lines correspond to the local density of states in three different columns in one unit cell. The energy of magnon is $\varepsilon=1.5$ in the lower bulk band gap. }
\end{figure}
Based on the formula Eq.~(\ref{seq_localj}), we can calculate the equilibrium or nonequilibrium magnon transport in the lattice.
Fig.~\ref{fig6_localeq} shows the edge state magnon transport in the bulk gap at a fixed magnon energy $\varepsilon=1.5$ in the thermal equilibrium.  The forward (left-to-right) thermal current carried by magnons travels along one edge, and the backward (right-to-left) current with the same magnitude transports along the other one, as shown in Fig.~\ref{fig6_localeq} (a). Near both edges the local magnon currents form many chiral vortices  due to the nonzero DM interaction. Moreover, both the current and the magnon density of states are symmetrically localized at two edges. We plot the local current and the local density of states for the edge mode with $\varepsilon=1.5$ in Fig.~\ref{fig7_jroy}. We find both the local current and the local density of states decay exponentially from the edge to the center with some oscillations. The oscillations come from the vortex of the energy current in the kagome lattice. Thus the magnon with energy  $\varepsilon=1.5$ in the lower bulk band gap indeed localizes at the two edges of the quasi-1D lattice. The magnon with other energies in the bulk gaps also has the similar picture.

When a defect is present at one edge, the current take a detour around it and transports ahead without backscattering, as illustrated in Fig.~\ref{fig6_localeq} (b). Although the defect dramatically affects the local density of magnons and destroys the local current vortex, the global currents along two edges keep intact compared to those of the uniform lattice, and their summation keeps zero since the net transport vanishes at equilibrium.

\begin{figure}[t]
\includegraphics[width=1.0\columnwidth]{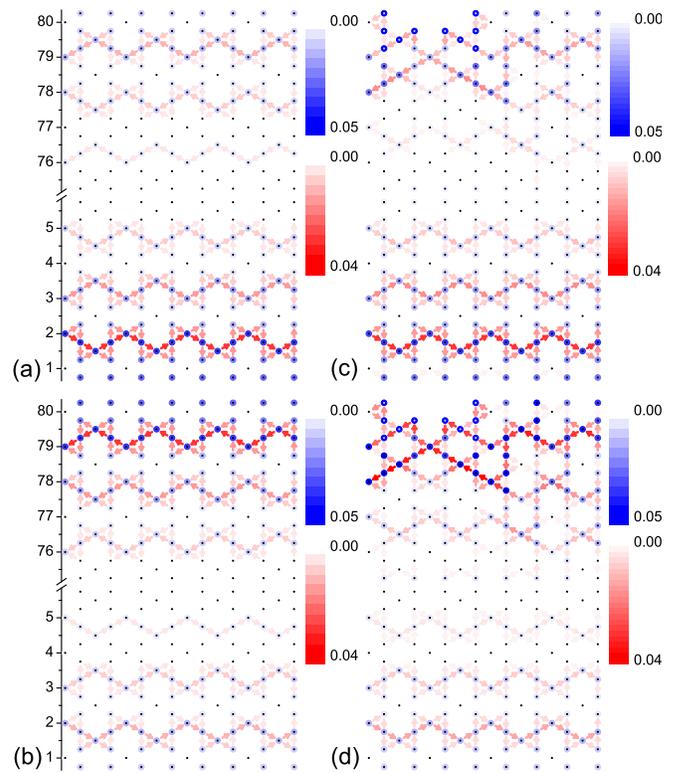}%
\caption{\label{fig8_localneq} (color online). The local energy current and density of state for edge magnon transport at nonequilibrium. The red arrows, the blue dots, and the small black dots correspond to the local energy current, the local density of magnon, and atom sites, respectively. The color of the arrows and dots indicate the magnitude of the local current and density of states, respectively. The uniform kagome lattice are with (a) $T_L=1.2, T_R=0.8$, and (b) $T_L=0.8, T_R=1.2$. The lattice with a defect at the upmost site of the left fifth column are with (c) $T_L=1.2, T_R=0.8$, and (d) $T_L=0.8, T_R=1.2$. Other parameters are $\varepsilon=1.5$, $D/J=0.1$, $a=1$, $W=80$.}
\end{figure}
As shown in Fig.~\ref{fig8_localneq} (a) and (b), when two leads are held at different temperatures, the magnons and energy current prefer to flow along one edge from the hot lead to the cold one. The transport around the other edge however shows an interesting anomalous behavior that the magnons and energy reversely flow from the cold lead to the hot one. Nevertheless, we note that this does not violate the second law of thermodynamics because the forward (hot-to-cold) energy current transported along one edge is larger than the backward one (cold-to-hot) along the other edge so that the total transport is still from the hot part to the cold one. If we only swap two temperatures ($T_L\leftrightarrow T_R$), the transport will prefer the other edge but with the directions of local edge currents unchanged. If we merely reverse the DM interaction ($D\rightarrow-D$), the transport will change to prefer the other edge with the local currents reversed but with the total average current unchanged. If we swap both the temperatures and the DM interaction, the local currents will just reverse their directions but keep the same magnitudes, which is a consequence of the time-reversal invariance.


It is worthy to notice that for the chiral edge state, although both the current and the magnon density of states mainly localize around two edges, they leak into the bulk with oscillatory decay. The oscillatory motion results from the quantum interference due to the edge boundaries, which is similar to the properties of the localized edge phonon modes \cite{jiang09} and electron transport in graphene \cite{qiao10}. This phenomenon indicates that even for the topological chiral edge state, the transport within the bulk of a topological insulator can not be really avoided. A topological insulator is not a perfect ``insulator'', not only referring to the edges/surfaces, but also for the bulk.

When a defect is present around one edge, the one-way edge current in the TMI is able to take a detour around it and transport ahead without backscattering, see Fig.~\ref{fig8_localneq} (c) and (d). Although the defect dramatically affects the local density of magnons and the vortex pattern of local currents, the global currents along two edges keep intact compared to those of the uniform lattice. This means that the chiral edge state in the bulk gap is indeed topologically protected from the lattice defect or weak disorders.

\section{  Thin film of  Lu$_2$V$_2$O$_7$}\label{secreal}
\begin{figure}
\includegraphics[width=1.00\columnwidth]{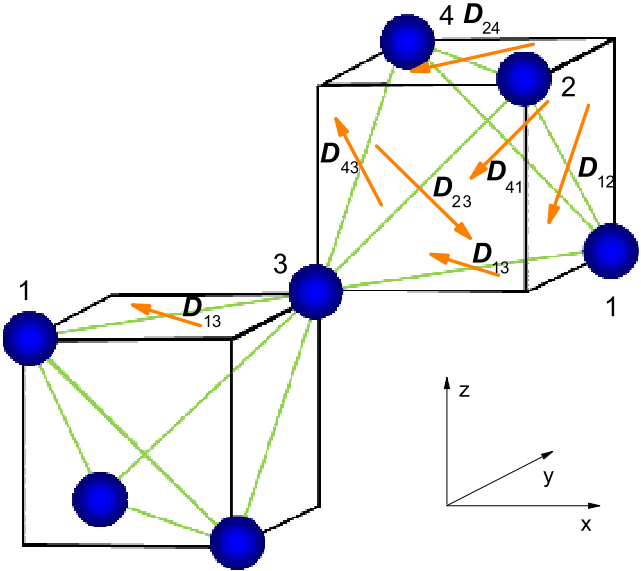}%
\vspace{-5mm}
\caption{\label{fig9_unit} Two tetrahedrons in the pyrochlore lattice of the atom vanadium of the ferromagnet Lu$_2$V$_2$O$_7$.  On a single tetrahedron, all the DM vectors on bonds 1-2, 2-4, 4-1, 1-3, 2-3, and 4-3 are shown by the arrows. }
\end{figure}

\begin{figure}[t]
\includegraphics[width=1.00\columnwidth]{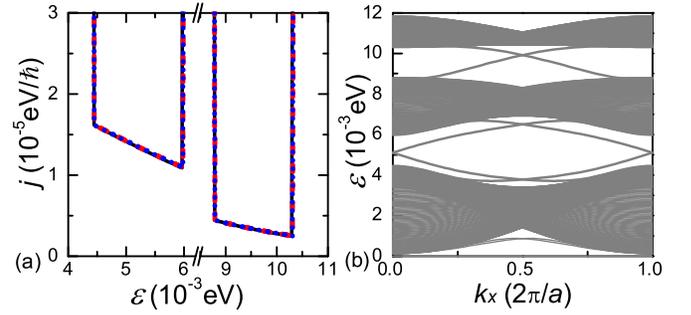}%
\vspace{-3mm}
\caption{\label{fig10_current} (color online). (a) The current density vs energy of magnon for uniform and edge-defect kagome lattices with the parameters of Lu$_2$V$_2$O$_7$. The solid and dotted lines correspond to the energy current in the bulk band gaps for uniform and edge-defect lattices, respectively.  (b) The dispersion relation of the kagome lattice with the parameters of Lu$_2$V$_2$O$_7$. $J=3.4$ {\rm meV}, $D/J=-0.26$, $H_{
0}=1$ T, $T_L=21$ K, and $T_R=19$ K. }
\end{figure}

In the insulating ferromagnet Lu$_2$V$_2$O$_7$, the orbitals of the $d$ electrons are ordered to point to the center of mass of the vanadium tetrahedron and a virtual hopping process stabilizes the ferromagnetic order of the vanadium spin in this orbital-ordered state \cite{onose10, ichikawa05}. The vanadium sublattice in Lu$_2$V$_2$O$_7$ has a pyrochlore structure composed of corner-sharing tetrahedra,  that is, a stacking of alternating  kagome and triangular lattices along the [111] direction.  Considering the strong constraint of the crystal symmetry and using Moriya's rules \cite{moriya60}, possible DM (Dzyaloshinskii-Moriya) interactions on a single tetrahedron can be determined as \cite{onose10}
$  \vec D_{12}  = {D_0 \over {\sqrt 2 }}( - \hat e_y  - \hat e_z ),\, \vec D_{24}  = {D_0 \over {\sqrt 2 }}( - \hat e_x  - \hat e_y ),\, \vec D_{41}  = {D_0 \over {\sqrt 2 }}( - \hat e_x  - \hat e_z ),\, \vec D_{13}  = {D_0 \over {\sqrt 2 }}( - \hat e_x  + \hat e_y ),\, \vec D_{23}  = {D_0 \over {\sqrt 2 }}( + \hat e_x  - \hat e_z ),\, \vec D_{43}  = {D_0 \over {\sqrt 2 }}( - \hat e_y  + \hat e_z )$.
Here $D_0$ denotes the strength of the DM interaction; the number 1,2,3, and 4 denote the site in  a single tetrahedron. If we apply a magnetic field $ \vec H_0 $, then all the spin angular momentum in the direction along $\vec l= \vec H_0 /H_0$, with $H_0$ the magnitude of $\vec H_0$. We know the component of the DM vector perpendicular to $\vec l$ does not contribute to the
spin-wave Hamiltonian, thus we only retain the projections of the DM interaction along $\vec l$ direction, i.e., $D_{mn}^l=\vec{D}_{mn}\cdot \vec{l}$.
If we apply a magnetic field along $\vec l= [111]$  direction, then $D_{13}^l=D_{23}^l=D_{43}^l=0$ and $D_{12}^l=D_{24}^l=D_{41}^l=\frac{-\sqrt{2} }{\sqrt 3 } D_0$ \cite{explanD}. Therefore, if the magnetic field is applied along $[111]$ direction, the magnon Hall effect and topological magnon insulator effect only come from the noncancellation of different types of DM interaction loops in the unit cell of the kagome lattice. The effective DM interactions between inter-layer sites have no contributions. According to the experimental observation in Ref.~\cite{onose10}, the DM interaction is obtained $D_0/J = 0.32$, thus we use $D={-\sqrt{2} \over {\sqrt 3 }}D_0={-\sqrt{2} \over {\sqrt 3 }}\cdot 0.32 J = -0.26 J$ for the DM interaction of the thin film of Lu$_2$V$_2$O$_7$ with kagome layer.  Since in Ref. \cite{onose10}, we have $JS=8D_s/a_0^2$ with $a_0=9.94 {\rm\AA}$ the spacing between unit cells of the pyrochlore structure and $D_s=21{\rm meV} {\rm \AA}^2$ the spin stiffness constant, then we get the coupling $J=3.4$ meV and $a=\frac{\sqrt2}{2}a_0=7.03$ {\AA}.  Based on these parameters, we calculate the dispersion relations for the quasi-1D kagome lattice, as shown in Fig.~\ref{fig10_current} (b).

As shown in Fig. \ref{fig10_current} (a), the energy current of magnon is not affected by defect or disorder in the range of $[4.45, 5.98]\cup[8.79, 10.31]$ meV. These energy intervals coincide with the bulk gaps in the magnon spectrum where the topological magnonic edge states can be identified [see Fig.~\ref{fig10_current} (b)].  Although certain distortions of edge states will occur as results of the defect or disorder, the total energy current carried by edge magnons does not change in the whole bulk energy gap. This indicates that the defect or disorder does not open a gap in the magnon spectrum so that the topology of the chiral magnon edge state is robust. According to the energy ranges, the topological magnon states have frequencies within $[1.08, 1.45]\cup [2.13, 2.49]$ THz.

Applying different external magnetic fields will not change the main properties of magnons, but shift the corresponding dispersion relations, so that the frequency of topological edge magnons can be tuned flexibly with a wide range. Also, when the inter-layer exchange couplings are considered, they only play the role of effective on-site potentials, which just shift the whole
bands and leave the main band structural properties unchanged. In addition, the two-dimensional kagome lattice sheet could be obtained by doping with nonmagnetic atoms as we mentioned before so that the inter-layer exchange couplings are ignorable. Therefore, we expect that one can observe the TMI for the thin film of Lu$_2$V$_2$O$_7$ in a wide energy range of magnons. Our findings about the TMI could also be applied for other magnetic crystals, including even antiferromagnetic materials where the existence of magnons is possible.

\section{Possible Experiments }
To realize magnonic devices as well as the predicted TMI, the excitation and detection of magnons is the major challenge. Recent years have witnessed a fast development in experimental techniques such as ferromagnetic resonance \cite{podbielski06}, pulse-inductive microwave magnetometer \cite{silva99}, time-resolved scanning Kerr microscopy \cite{barman03}, optical pump-probe techniques \cite{kimel07}, as well as Brillouin light scattering (BLS) \cite{perzlmaier05} which takes a special role since it allows the direct measure of dispersions and band structures. We could observe the topological edge modes by using these techniques to measure the magnon dispersion relation and could also verify the TMI by detecting the magnon transport in the bulk band gap where magnons could be selectively excited by non-thermal optical pulses \cite{kimel04,bigot09,zhangy12,kamp11,nish12} or induced by external spin-polarized current \cite{kiselev03}, so that we can avoid the thermal transport from bulk states but extract the one only from edge channels. We hope our theoretical predictions about TMI could open a new window into the application of nondissipative magnon transport, especially for the novel magnonic device design, which could also shed light on the information technology based on magnonics, and micro-spintronics.

\emph{Note added:} Recently, we have learned of a submission \cite{note} thanks to its authors, studying the similar topological chiral magnonic edge mode. Differently, their results are based on a linearized Landau-Lifshitz equation and account for the dipolar interaction instead of the DM one studied here.

\begin{acknowledgments}
L. Z. and B. L. are supported by the grant
R-144-000-285-646 from Ministry of Education of Republic of Singapore.
J. R. acknowledges the support of the U.S. Department of Energy through the LANL/LDRD Program.
J.-S. W. acknowledges support from a URC research grant R-144-000-257-112 of NUS.
\end{acknowledgments}


\begin{thebibliography}{99}
\bibitem{explanD} The projection of DM interaction along $\vec l= \vec H_0 /H_0$ can be written as
$
D_{12}^l  = {D_0 \over {\sqrt 2 }}{{ - H_y  - H_z } \over H_0},\, D_{24}^l  = {D_0 \over {\sqrt 2 }}{{ - H_x  - H_y } \over H_0},\, D_{41}^l  = {D_0 \over {\sqrt 2 }}{{ - H_x  - H_z } \over H_0},\,
D_{13}^l  = {D_0 \over {\sqrt 2 }}{{ - H_x  + H_y } \over H_0},\, D_{23}^l  = {D_0 \over {\sqrt 2 }}{{ + H_x  - H_z } \over H_0},\, D_{43}^l  = {D_0 \over {\sqrt 2 }}{{ - H_y  + H_z } \over H_0}
$. If the magnetic field is in the direction of $ [111]$, then we have $H_x=H_y=H_z=H_0/\sqrt3$.
\end{thebibliography}
\end{document}